\documentclass[runningheads]{llncs}
\usepackage{graphicx}
\usepackage[T1]{fontenc}
\usepackage{float}
\usepackage[hidelinks]{hyperref}
\usepackage{color}
\renewcommand\UrlFont{\color{blue}\rmfamily}
\usepackage[T1]{fontenc}
\usepackage{lmodern}

\DeclareUrlCommand{\tturl}{\urlstyle{tt}}
\DeclareUrlCommand{\bftturl}{\def\UrlFont{\bfseries}}
\DeclareUrlCommand{\bfurl}{\def\UrlFont{\bfseries\ttfamily}}
\newcommand{\hochkomma}{$^{,}$}

\usepackage{xcolor}
\DeclareUrlCommand{\redurl}{\def\UrlFont{\ttfamily\color{red}}}

\usepackage[normalem]{ulem}

\makeatletter
\newcommand*\DeclareFancyUrlCommand[2]{%
    \expandafter\DeclareFancy@UrlCommand
    \expandafter{\csname fancyurl@\expandafter\@gobble\string#2\endcsname}{#1}{#2}%
}
\def\DeclareFancy@UrlCommand#1#2#3{%
    \let#1\empty 
    \useunder{#3}{#1}{}%
    \DeclareUrlCommand{#2}{\def\UrlFont{\ttfamily#1}}%
}
\makeatother

\DeclareFancyUrlCommand{\ulurl}{\uline}
\DeclareFancyUrlCommand{\uulurl}{\uuline}
\DeclareFancyUrlCommand{\uwurl}{\uwave}
\DeclareFancyUrlCommand{\sourl}{\sout}
\DeclareFancyUrlCommand{\xourl}{\xout}
\DeclareFancyUrlCommand{\daurl}{\dashuline}
\DeclareFancyUrlCommand{\dourl}{\dotuline}

\usepackage{mdframed,color}
\usepackage{xcolor}
\usepackage{multirow}
\usepackage{subcaption}
\usepackage{listings}
\definecolor{codegreen}{rgb}{0.0,0.0,0.0}
\definecolor{codegray}{rgb}{0.5,0.5,0.5}
\definecolor{codepurple}{rgb}{0.58,0,0.82}
\definecolor{backcolour}{rgb}{0.97, 0.96, 1.0}

\begin{document}

\title{Caching HTTP 404 Responses Eliminates Unnecessary Archival Replay Requests}
\titlerunning{Caching HTTP 404 Responses Eliminates Unnecessary}
\author{Kritika Garg\inst{1}\orcidID{0000-0001-6498-7391} \and
Himarsha R. Jayanetti\inst{1}\orcidID{0000-0003-4748-9176} \and Sawood Alam\inst{2}\orcidID{0000-0002-8267-3326} \and Michele C. Weigle\inst{1}\orcidID{0000-0002-2787-7166} \and Michael L. Nelson\inst{1}\orcidID{0000-0003-3749-8116} }
\authorrunning{K. Garg et al.}
%
\institute{Old Dominion University, Norfolk, VA 23529, USA \\ \email{\{kgarg001,hjaya002\}@odu.edu} \email{\{mweigle,mln\}@cs.odu.edu} \and
Wayback Machine, Internet Archive, San Francisco, CA 94118, USA
\email{sawood@archive.org}\\}
\maketitle              
\begin{abstract}
Upon replay, JavaScript on archived web pages can generate recurring HTTP requests that lead to unnecessary traffic to the web archive. In one example, an archived page averaged more than 1000 requests per minute. These requests are not visible to the user, so if a user leaves such an archived page open in a browser tab, they would be unaware that their browser is continuing to generate traffic to the web archive. We found that web pages that require regular updates (e.g., radio playlists, updates for sports scores, image carousels) are more likely to make such recurring requests. If the resources requested by the web page are not archived, some web archives may attempt to patch the archive by requesting the resources from the live web. If the requested resources are not available on the live web, the resources cannot be archived, and the responses remain HTTP 404. Some archived pages continue to poll the server as frequently as they did on the live web, while some pages poll the server even more frequently if their requests return HTTP 404 responses, creating a high amount of unnecessary traffic. On a large scale, such web pages are effectively a denial of service attack on the web archive. Significant computational, network, and storage resources are required for web archives to archive and then successfully replay pages as they were on the live web, and these resources should not be spent on unnecessary HTTP traffic. Our proposed solution is to optimize archival replay using Cache-Control HTTP response headers. We implemented this approach in a test environment and cached HTTP 404 responses that prevented the browser's requests from reaching the web archive server.

\end{abstract}
\section{Introduction}
\label{introduction}
Web archives allow users to replay and browse archived web pages, or mementos, as they were on the live web. However, playback of the archived web pages may not be complete if the embedded resources are missing from the archive \cite{Brunelle2014missing}. We discovered that upon replay some archived web pages make recurring requests for missing embedded resources, creating unnecessary and wasteful traffic for the web archive. These recurring requests could only be seen by observing the network activity of the archived web page. To users browsing the web archive, these web pages would appear like any other regular mementos replaying in their browser. They would not see the web traffic generated by these recurring requests. Thus, if a user leaves such an archived page replaying in a browser tab for a long time, they would be unaware that their browser is generating a huge amount of unnecessary traffic to the web archive. Our previous work \cite{accesspattern2022tpdl} on analyzing the access logs of Arquivo.pt \cite{billionarch} revealed a similar user session. This long-running user session ran for almost four days and issued over 4.3M recurring requests to the web archive for the embedded images of a single memento of \texttt{radiocomercial.iol.pt}. We discuss this memento further in Section \ref{carousel_ex}. 

This discovery inspired us to investigate the kinds of web pages that would generate recurring requests to web archives similar to  \texttt{radiocomercial.iol.pt}. By examining the network traffic on numerous archived web pages, we found that web pages that require regular updates (e.g., sports scores updates, stock market updates, news updates, chat applications, new tweets) and poll the server periodically for the updates may generate recurring requests. For example, a popular and well-archived domain like \texttt{twitter.com} also exhibits this behavior when it polls for new tweets and the latest trends. The example of \texttt{twitter.com} is different from \texttt{radiocomercial.iol.pt} as it makes fewer recurring requests per minute. However, the cumulative load of many people globally replaying the mementos of \texttt{twitter.com} would result in a significant amount of wasted bandwidth for the web archive. We also saw that web pages with image carousels, banners, widgets, etc. are also more likely to cause the recurring requests. \looseness=-1

We studied the behavior of such web pages in different web archives.  Some web archives may patch the memento by requesting missing embedded resources from the live web. The patch/write requests would be successful only if the requested resources are accessible on the live web. However, if the requested resources are not accessible on the live web, the resources cannot be archived, and the patch/write requests would result in HTTP 404 responses. In this case, patching the memento from the live web would create unnecessary writes and reads. We describe a memento displaying this behavior in Section~\ref{playlist_ex2}. 

We found that some mementos would send requests to the server as often as they did on the live web, while others would poll the server even more rapidly if their requests returned HTTP 404 responses, resulting in excessive load on the web archive. On a large scale, web pages like these could effectively be the denial of service attacks, squandering network resources, overloading web archive servers, and possibly depriving other users access to the archive. Web archiving and archival replay are resource-intensive processes, and these resources should not be spent on unnecessary HTTP traffic. That is why it is important to be aware of such issues and optimize the replay system accordingly for an effective playback. Eliminating this wasteful HTTP traffic to the web archives will also have a positive, although small, environmental impact. In this paper, we describe various sources that could cause unnecessary HTTP traffic for the web archives. In Section~\ref{things_we_found_in_wild}, we provide examples of the mementos that generate unnecessary recurring requests. In Section~\ref{abstract_model}, we demonstrate a minimal reproducible example web page containing a carousel that generates recurring requests for the missing embedded resources. In Section~\ref{solution}, we implement a solution for eliminating unnecessary requests by using the Cache-Control HTTP response header to cache HTTP 404 responses.

\section{Background and Related Work}
\label{background}

Web archiving involves recording HTTP traffic from web servers and then replaying them in a different context. A memento, or URI-M, is a snapshot of a URI-R (URI of an original resource) captured at a specific Memento-Datetime (the datetime a particular URI-M was archived). These terms are defined in the Memento Protocol RFC \cite{memento:rfc}.

The objective of successful archival replay is that when replaying an archived web page, the page should be viewable and behave exactly as it did at the time of archiving. To render a web page the way it looked in the past, the base HTML page and all the related embedded resources, such as images, stylesheets, JavaScript, fonts, and other media, should be archived around the same time as the base page. However, not every embedded resource of the page that is attempted to be archived is captured by web archives. As a result, some of the embedded resources on archived pages are missing. Brunelle et al. \cite{Brunelle2014missing} have measured the impact of missing resources in web archives. The missing embedded resources may introduce anomalies during archival replay.

For example, in our study, we saw that various mementos repeatedly made requests to missing resources during the replay, causing unnecessary or wasteful traffic for the web archive. In our previous work \cite{Garg2021Twitter}, we documented the difficulties in replaying mementos of Twitter’s new user interface due to missing embedded JSON files. Missing resources could also lead to temporal discrepancies during replay. In the Internet Archive's Wayback Machine, the Memento-Datetime of the base HTML page and the Memento-Datetime of the corresponding embedded resources may or may not be temporally aligned, which could result in a temporal violation during the replay  \cite{ainsworth2014framework,ainsworth2015only}. 

These anomalies can cause security vulnerabilities in web archives. As an example, these unnecessary recurring requests on a wide scale may overwhelm a web archive with excessive web traffic, leading to the denial of archival services. Additional research into the security of high-fidelity web archives  \cite{iilyacushman,meltingpot-watanabe} has revealed a number of security risks to web rehosting services. Lerner et al. \cite{rewritehistory-adalerner} detected several vulnerabilities and security attacks specific to the Internet Archive’s Wayback Machine. The security issues raised above show the importance of optimizing and upgrading the archival replay systems. Goel et al. \cite{goel2022jawa} proposed a design to reduce storage needs by discarding JavaScript code with functionality that will not work or would remain unexecuted during replay. Our work focuses on the effects of JavaScript code that executes during replay and triggers recurring HTTP requests. Our proposed solution does not cause an overhaul of the system because we add the header to the server layer without changing the application. We can eliminate the wasteful network traffic caused by the executing JavaScript code by returning a Cache-Control HTTP header.  

\section{Things We Found in the Archive}
\label{things_we_found_in_wild}
We examined several mementos that cause recurring requests upon replay. We noticed this behavior in mementos with missing resources for the banners, widgets, carousels, playlists, and web pages that request regular updates (e.g., updates for sports scores). 
In this section, we provide examples of five such mementos. 

\subsection{Banner Example}\label{banner_ex}
\begin{figure}[ht]
\centering
\includegraphics[width=1\textwidth]{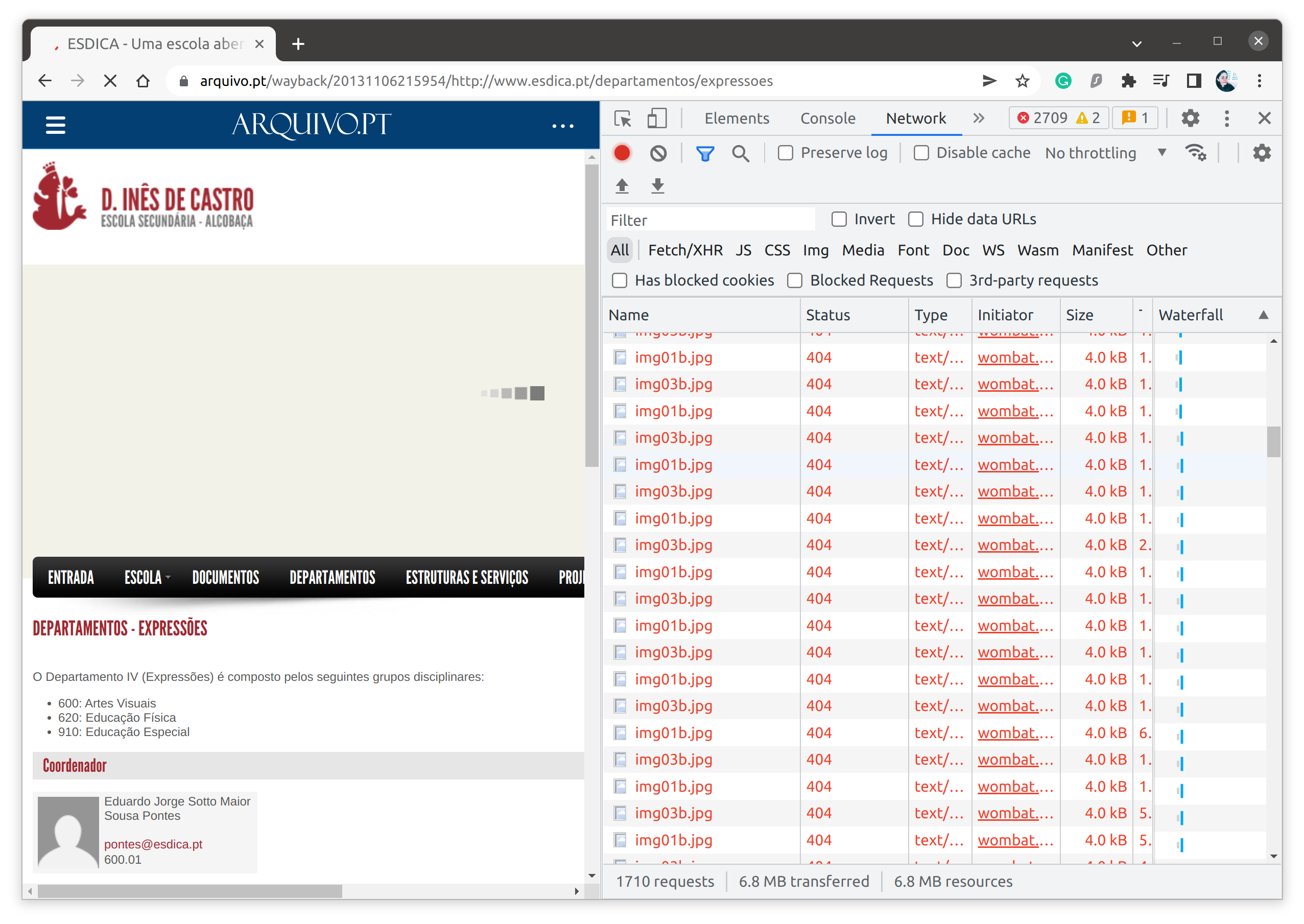}
\caption{Memento of a school website with a banner making recurrent requests for missing embedded images to Arquivo.pt   \url{https://arquivo.pt/wayback/20131105211447/http://esdica.pt/}}  
\label{fig:banner_img}
\end{figure}

Figure~\ref{fig:banner_img} shows a memento of  \url{http://esdica.pt/} captured on 2013-11-06T21:59:54 in  Arquivo.pt and its network activity in Chrome DevTools \cite{Chrometools}. The memento is of the homepage of a high school website that contains a large banner trying to display a series of images in the form of a slideshow. The banner slideshow is generated by a jQuery Advanced Slider component that cycles through a list of images in an endless loop.\footnote{\url{https://arquivo.pt/wayback/20131105212033js_/http://esdica.pt/js/slider/jquery.advancedSlider.min.js}} The network tab of Chrome DevTools shows that the HTTP GET requests for the embedded images received HTTP 404 Not Found responses from the web archive. This means that the requested mementos of the embedded images are not available in the web archive. We noticed that this memento is making recurring HTTP GET requests to Arquivo.pt for the missing images.\footnote{\url{https://arquivo.pt/wayback/20131105211447/http://esdica.pt/imagens/banners/img03b.jpg}} 
This banner is shared across many pages at \texttt{esdica.pt}, which means all the archived pages would generate similar loads.\looseness=-1


\subsection{Carousel Example}\label{carousel_ex}
\begin{figure}[h]
\centering
\includegraphics[width=1\textwidth]{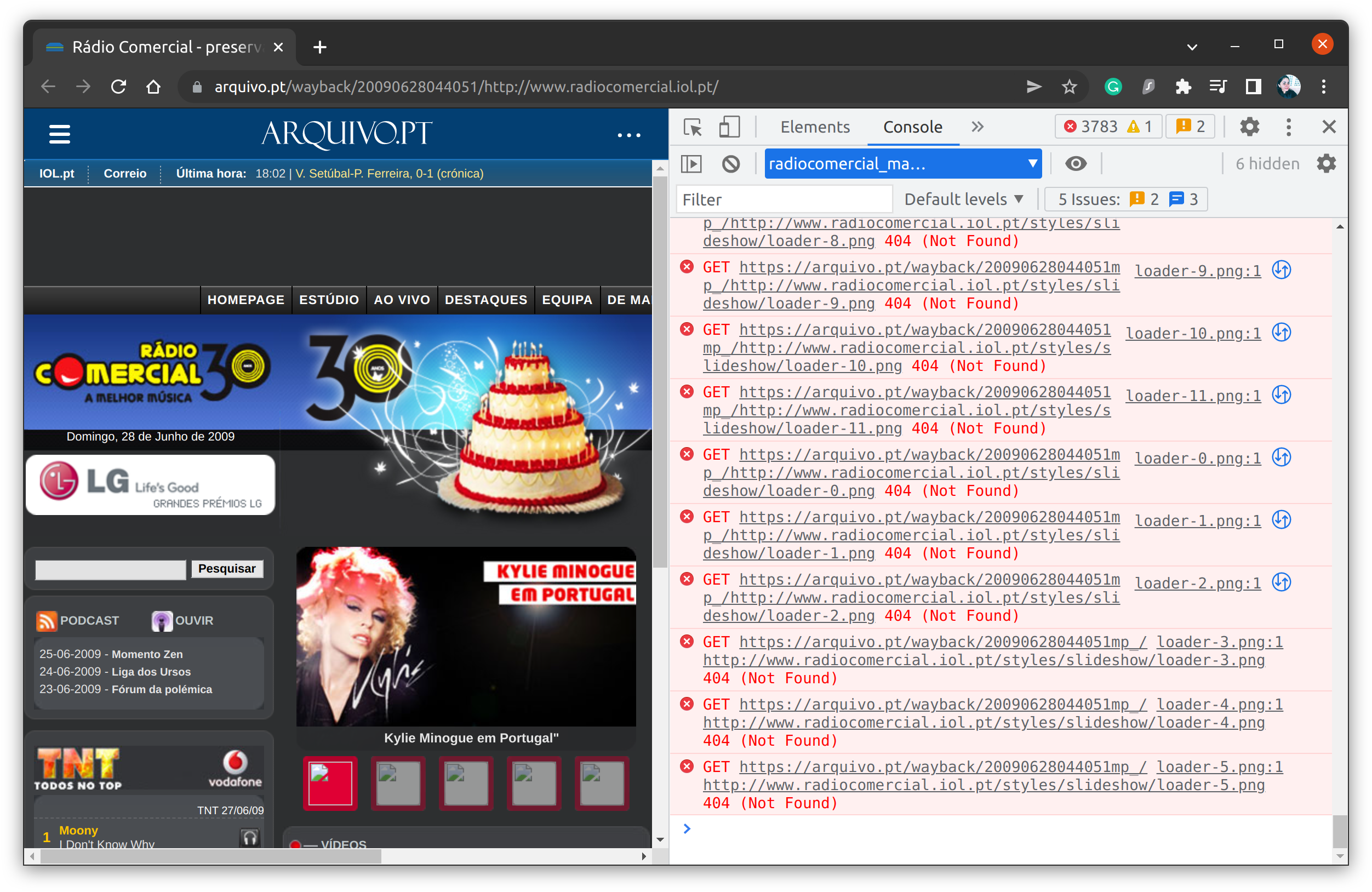}
\caption{Memento of a radio station web page with a carousel making recurrent requests for missing embedded images to Arquivo.pt   \url{https://arquivo.pt/wayback/20090628044051/http://www.radiocomercial.iol.pt/}} 
\label{fig:carousel_img}
\end{figure}

Figure~\ref{fig:carousel_img} shows a memento of \url{http://www.radiocomercial.iol.pt/} captured on 2009-06-28T04:40:51 in Arquivo.pt. Upon replay, it makes recurring HTTP GET requests for the embedded images\footnote{\url{https://arquivo.pt/wayback/20090628044051im_/http://www.radiocomercial.iol.pt/styles/slideshow/loader-0.png}} to the web archive server.
This memento of the Rádio Comercial website contains a carousel with a slideshow cycling through a series of images of musicians. The carousel is built with Cascading Style Sheets (CSS) and JavaScript.\footnote{\url{https://arquivo.pt/wayback/20090628052553js_/http://www.radiocomercial.iol.pt/jscript/slideshow/slideshow.js}} The JavaScript contains a loader function that iterates through a series of 12 images in the form of \texttt{/styles/slideshow/loader-\#.png}, where \texttt{\#} is replaced by numbers from 0 to 11 and loads them for the slideshow. If an image is not available, then the next time through the slideshow, another GET request will be made for the image. The Network tab in Figure~\ref{fig:carousel_img} shows that the HTTP GET requests for the images received HTTP 404 Not Found responses from the web archive. Therefore, JavaScript keeps sending requests for the embedded images, as it cannot load them for the slideshow. This results in unnecessary traffic to the web archive server. We observed that replaying this memento causes 122,204 requests in 10 minutes (1098.36 requests per minute on average) to the web archive.

\subsection{Playlist Example 1}\label{playlist_ex1}
\begin{figure}[h]
\centering
\includegraphics[width=1\textwidth]{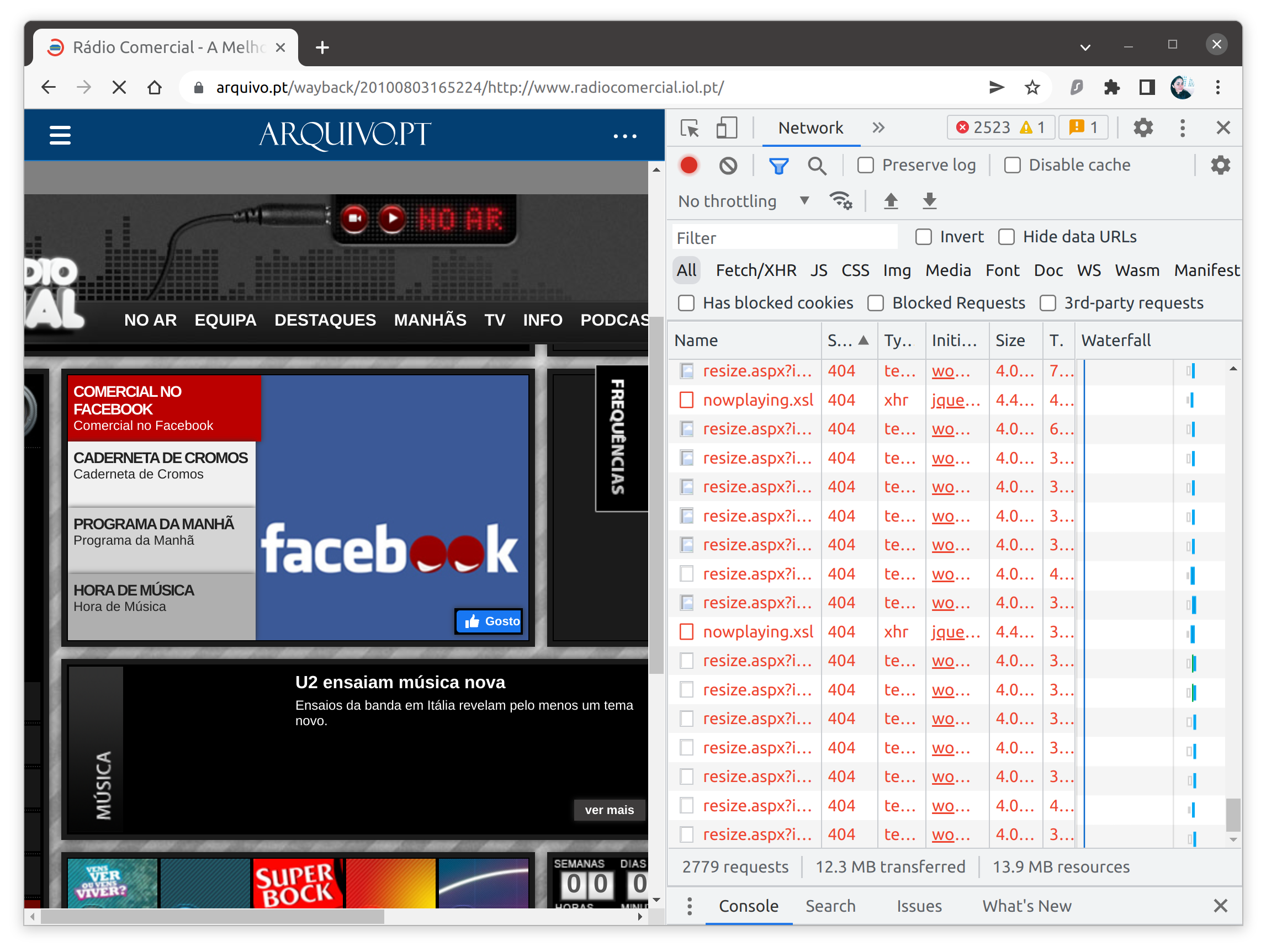}
\caption{Memento of radio webpage with playlist making recurrent requests for missing embedded images to Arquivo.pt.   \url{https://arquivo.pt/wayback/20100803165224/http://www.radiocomercial.iol.pt/}} 
\label{fig:playlist1_img}
\end{figure}

Figure~\ref{fig:playlist1_img} shows another memento of \url{http://www.radiocomercial.iol.pt/} captured on 2010-08-31T16-52-24 in Arquivo.pt, a year after the memento described in Section \ref{carousel_ex}. Between the two captures, the Rádio Comercial website completely changed its user interface. However, the behavior of generating recurring HTTP GET requests persisted. We observed that there were two forms of requests that were recurring.\footnote{\url{https://arquivo.pt/wayback/20100803165224mp_/http://www.radiocomercial.iol.pt/global_aspx/resize.aspx}}\hochkomma\footnote{\url{https://arquivo.pt/wayback/20100803165224mp_/http://www.radiocomercial.iol.pt/xsl_files/includes/nowplaying.xsl}}

The memento contains a playlist of songs and cover images for the songs. Figure~\ref{fig:playlist1_snip} shows the snippet of the source code of the memento, showing how the images were loaded for the playlist. The code contains the \texttt{onerror} property, which processes error events. When the image resource fails to load due to missing images in the archive, the \texttt{call\_resize} function is initiated. This function generates the request for the images from a different source, \url{http://www.radiocomercial.iol.pt/global_aspx/resize.aspx}. Requests for the images from this new source also received HTTP 404 responses. We saw recurring requests for this resource while observing the network activity of the memento in Chrome DevTools. The other recurring request for \url{http://www.radiocomercial.iol.pt/xsl_files/includes/nowplaying.xsl} was made via the XMLHttpRequest (XHR) object. The memento contains external JavaScript that is responsible for generating these requests. All of these XHR requests also received HTTP 404 responses from the server.

\begin{figure}[ht]
\begin{lstlisting}[numbers=none, backgroundcolor = \color{white}]

<img style="width: 35px; height: 35px; border:1px solid #fff; vertical-align: middle; margin: 1px 4px 0px 4px; float: left;" 
<@\textcolor{red}{src=}@> "/web/20100822133654im_/http://www.radiocomercial.iol.pt/global_aspx/images/o_aprendiz_de_feiticeiro_300[{B}upload{B}O{B}][35X35].jpg" 
<@\textcolor{red}{onerror}@>="call_resize (this, '/upload/O/o_aprendiz_de_feiticeiro_300.jpg', 35, 35);">}

function call_resize(_1,_2,_3,_4)}{_1.src="/global_aspx/resize.aspx?img="+_2+"&h="+_3+"&w="+_4;};

\end{lstlisting}
 \caption{A snippet of the source code of the memento of a radio webpage initiating the request for images}
\label{fig:playlist1_snip}
\end{figure}

\subsection{Playlist Example 2}\label{playlist_ex2}

\begin{figure}[!ht]
\centering
\includegraphics[width=1\textwidth]{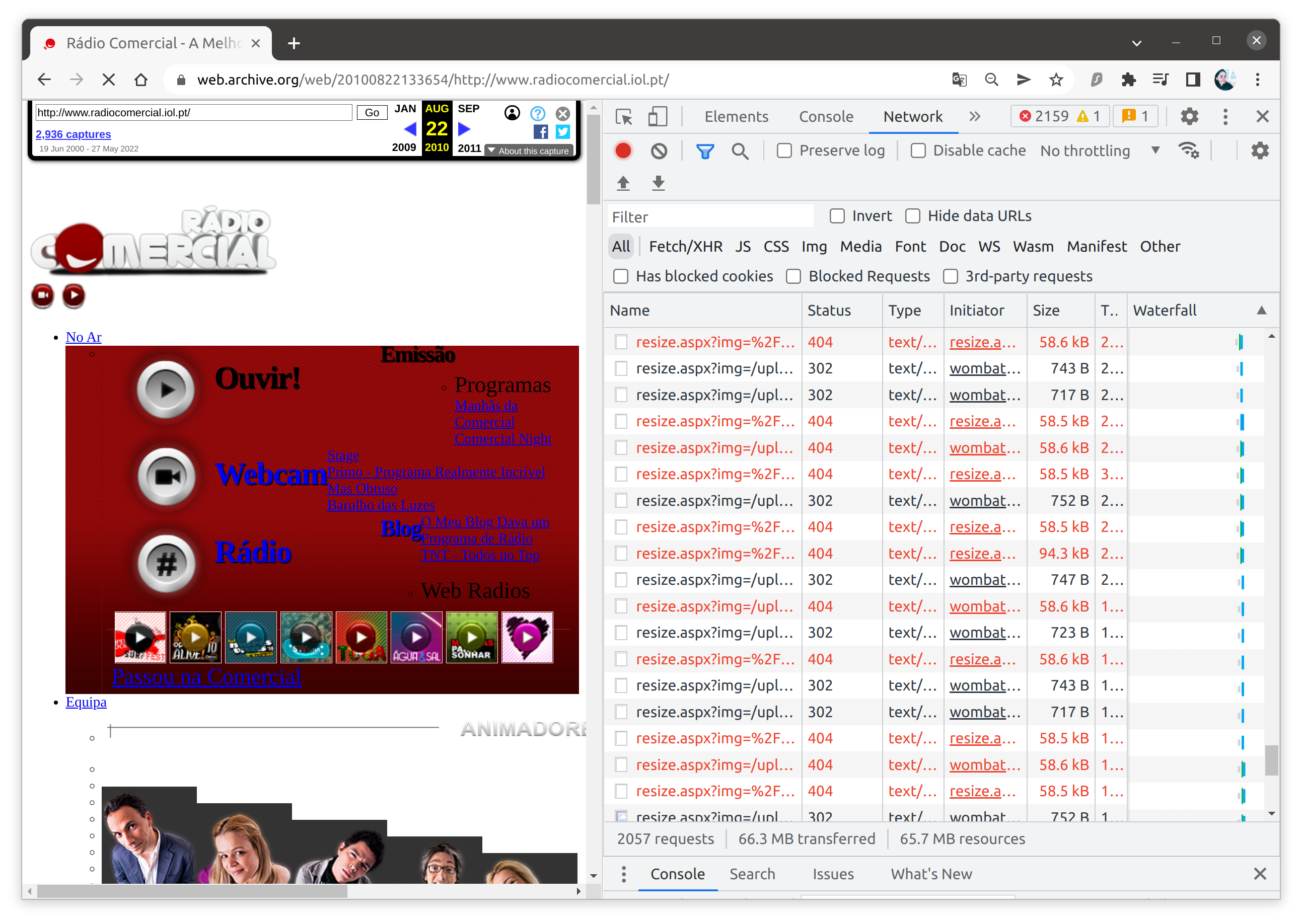}
\caption{Memento of a radio webpage with a playlist making recurrent requests for missing embedded images to the Internet Archive web server \url{https://web.archive.org/web/20100822133654/http://www.radiocomercial.iol.pt/}} 
\label{fig:playlist2_img}
\end{figure}

We observed a memento of the same URI-R as in Section~\ref{playlist_ex1} captured by the Internet Archive (IA) a few days before Arquivo.pt memento. Figure~\ref{fig:playlist2_img} shows another memento of \url{http://www.radiocomercial.iol.pt/} captured on 2010-08-22T13:36:54 in the Internet Archive’s Wayback Machine. This allowed us to observe how different web archives handle such mementos.

The memento contains a playlist of songs and cover images for the songs. We identified that the memento captured by the  Internet Archive (IA) was missing many embedded resources such as CSS, JavaScript, images, etc. The memento was not rendered correctly due to missing CSS. We inspected the network activity in Chrome DevTools and noticed that the memento was also making the same recurring GET requests and receiving HTTP 404 responses for the images in the playlist (Figure~\ref{fig:playlist2_img}).

\begin{figure}[ht]
\centering
\includegraphics[width=1\textwidth]{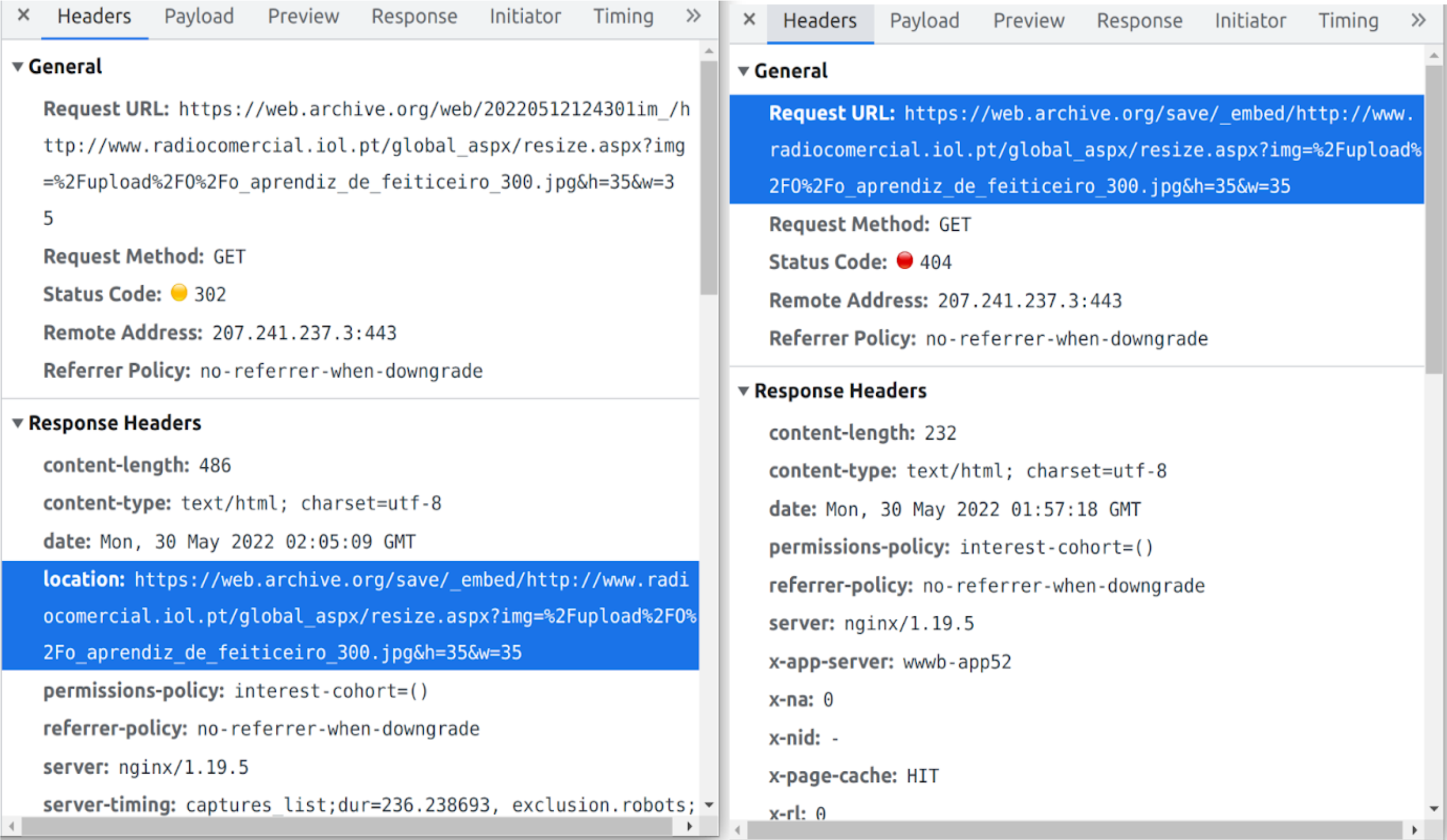}
\caption{Request for the missing resource is redirected to IA's SPN service to archive the resource (left), and the request to archive the missing resource returned HTTP 404 response (right)} 
\label{fig:playlist2_spn}
\end{figure}

IA tries to patch mementos by requesting the missing embedded resources from the live web to archive them during the replay. This is done with their Save Page Now (SPN) service \cite{SPN-mark}, which issues a request in the form \texttt{https://web.archive.org/\textcolor{red}{save/\_embed}/\{URL\}} (Figure~\ref{fig:playlist2_spn}, left). The request to patch the missing resource received an HTTP 404 response, indicating that the image does not exist on the live web (Figure~\ref{fig:playlist2_spn}, right). This resulted in multiple recurring requests for the same resource, recurring read requests for the memento, and recurring SPN requests (via \texttt{/save/\_embed/}) for the memento to the IA web server. We observed that, on average, 30 seconds after the first request, the new SPN requests receive HTTP 429 Too Many Requests responses from IA in an effort to throttle the excessive number of SPN requests made to its server. \sloppy 

\subsection{Latest Feed Example}\label{live_ex}
\begin{figure}[h]
\centering
\includegraphics[width=1\textwidth]{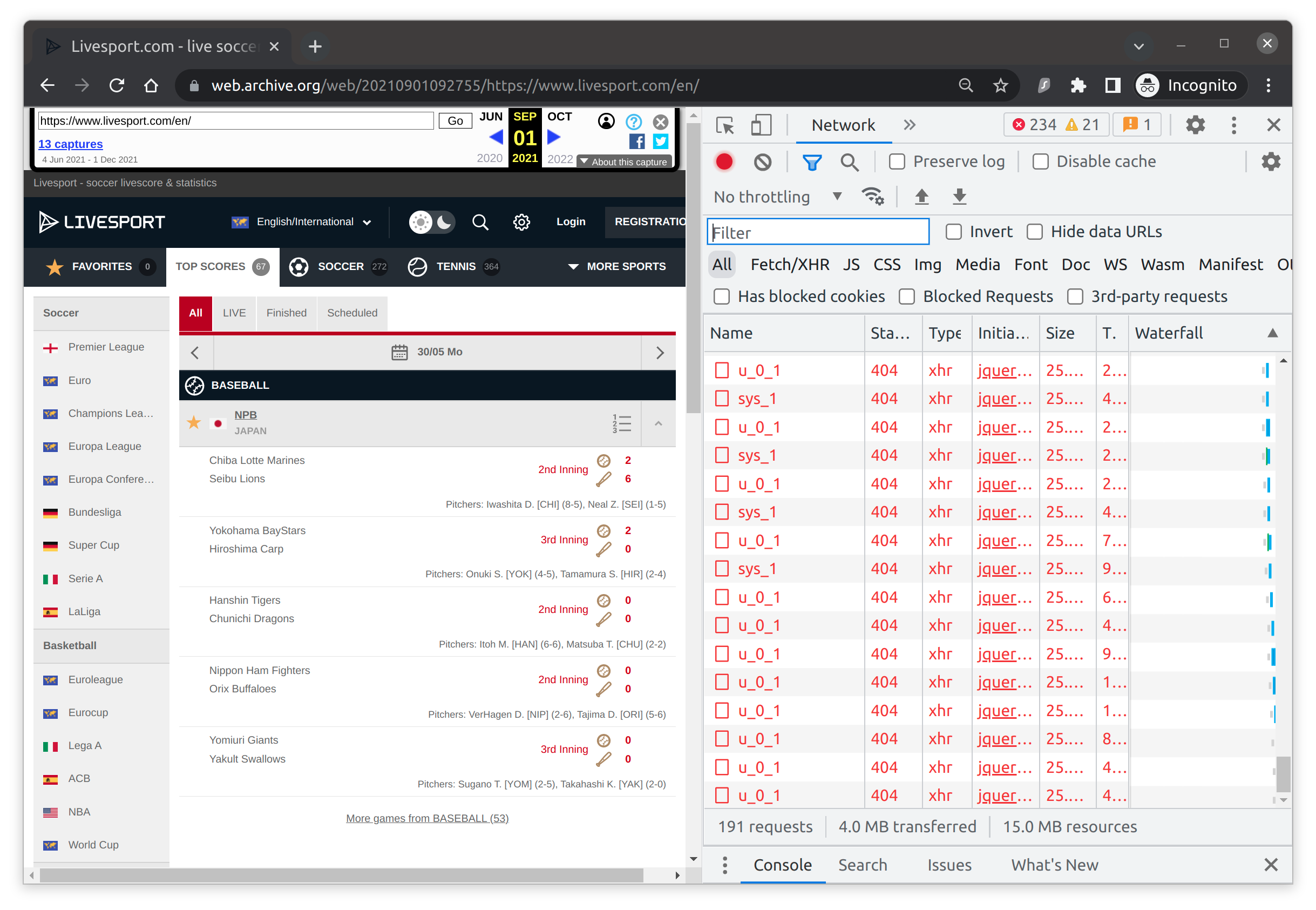}
\caption{Memento making recurrent XHR requests to fetch the live score API feeds   \url{https://web.archive.org/web/20210901092755/https://www.livesport.com/en/}
} 
\label{fig:livefeed_img}
\end{figure}

Figure~\ref{fig:livefeed_img} shows the memento of \url{https://www.livesport.com/en/} captured on 2021-09-01T09:27:55 in the Internet Archive’s Wayback Machine. The memento is of a sports website that provides live score updates for different sports. When the memento is loaded, the web page tries to fetch the memento of the scores API feed. The XMLHttpRequest (XHR) request returned an HTTP 404 response since the feed is not archived. The web archive cannot archive the feeds because it requires authorization. The memento keeps requesting the feed resulting in recurring unsuccessful requests. For example, the memento in Figure~\ref{fig:livefeed_img} is making recurring requests for two feeds.\footnote{\url{https://web.archive.org/web/20210901092756/https://d.livesport.com/en/x/feed/u_0_1}}\hochkomma\footnote{\url{https://web.archive.org/web/20210901092756/https://d.livesport.com/en/x/feed/sys_1}}

\section{Abstract Model: Minimal Reproducible Example}
\label{abstract_model}
In our examples, web pages that require regular feed updates or contain carousels, banners, widgets, etc. are more likely to make these recurring requests, causing a surge in web traffic. We implemented a minimal reproducible example (MRE) web page\footnote{\url{https://kritikagarg.github.io/Unnecessary-Archival-Replay-Requests/MREcarousel.html}} with a carousel to assess this behavior in a simpler environment. 

\begin{figure}[ht]
\centering
\includegraphics[width=1\textwidth]{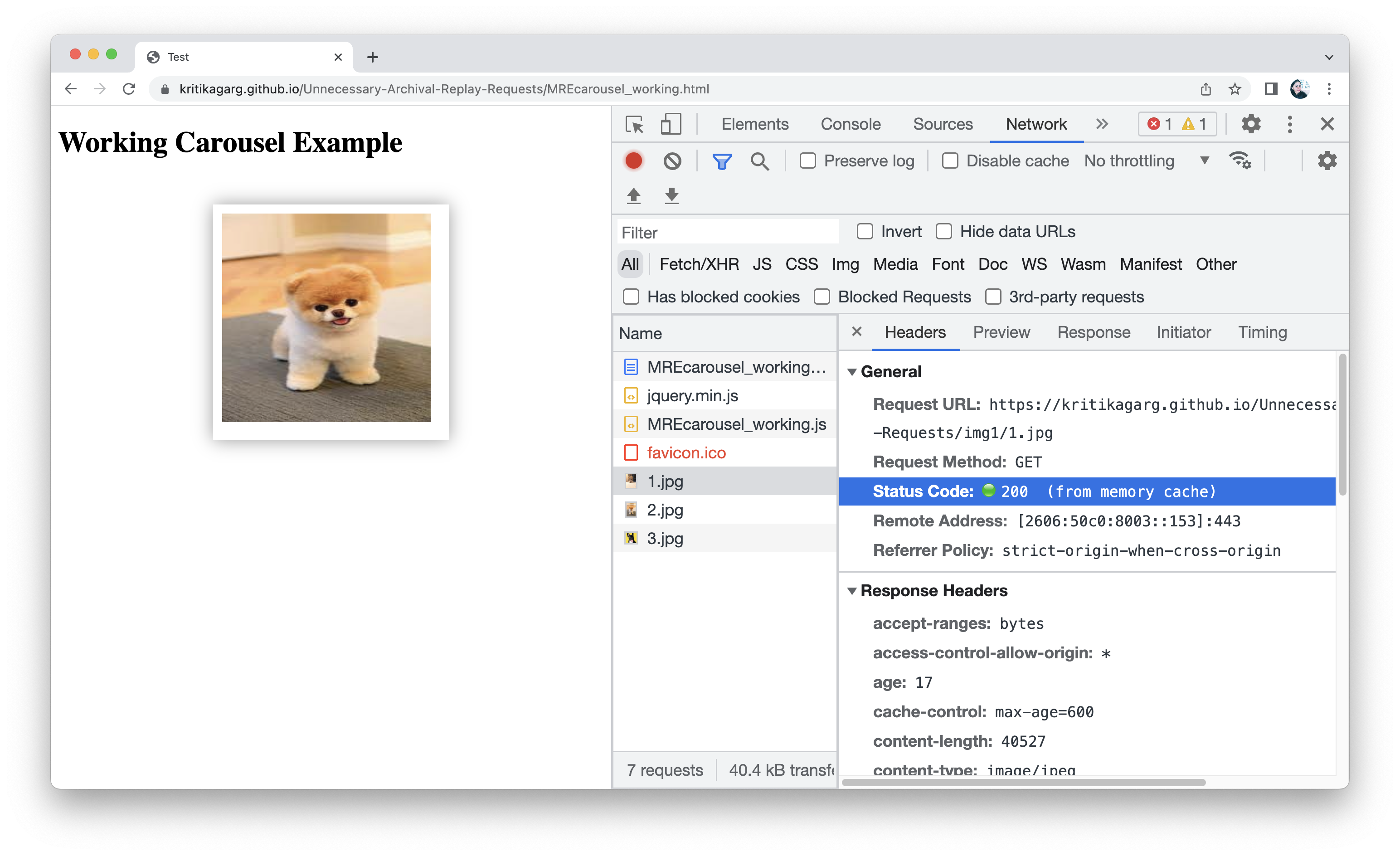}
\caption{MRE carousel serving requested embedded resources from the memory cache (request in Chrome DevTools) \url{https://kritikagarg.github.io/Unnecessary-Archival-Replay-Requests/MREcarousel_working.html}}
\label{fig:mre_working}
\end{figure}

\begin{figure}[ht]
\centering
\includegraphics[width=1\textwidth]{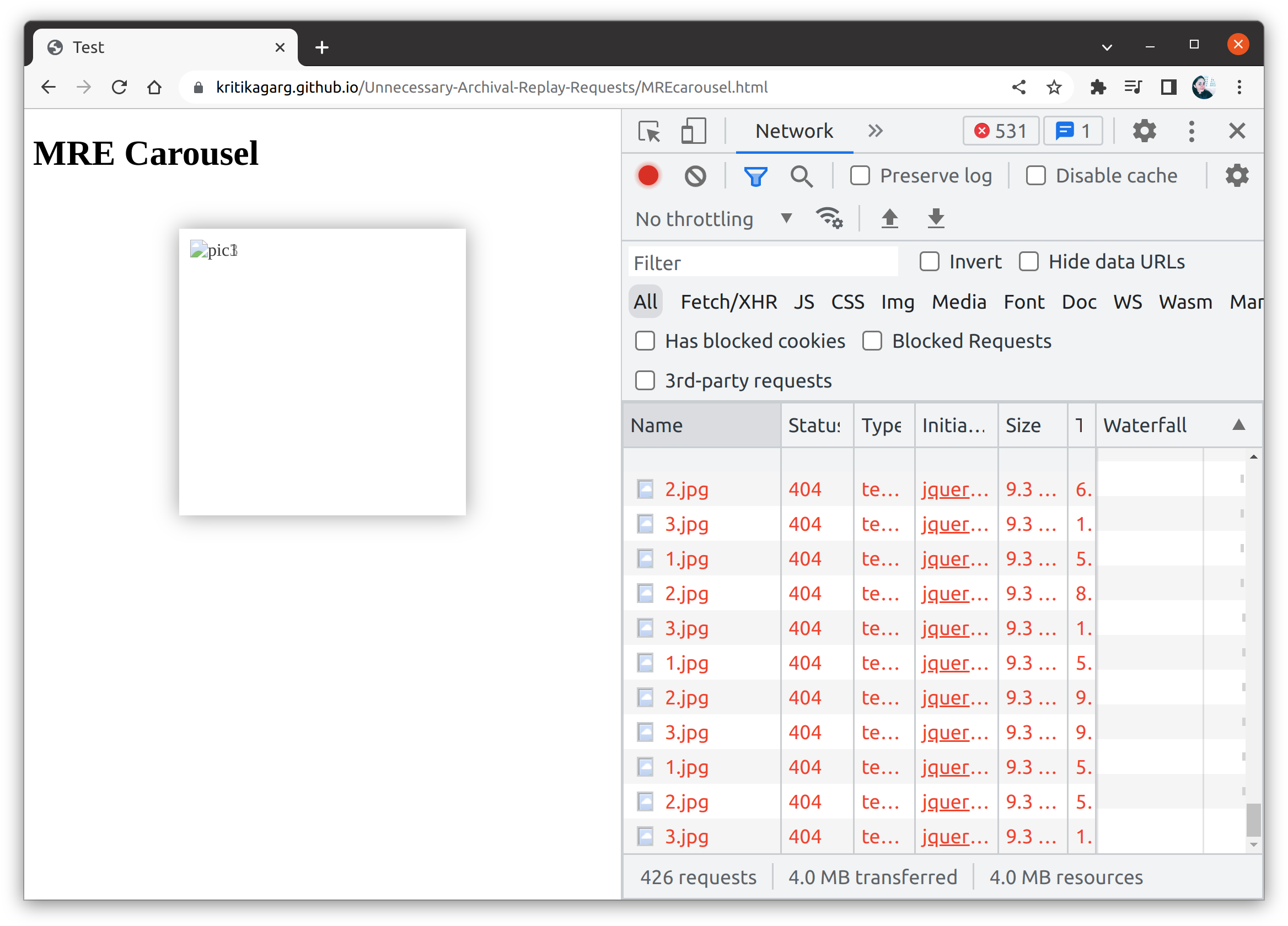}
\caption{MRE carousel making recurring requests for missing embedded resources (request in Chrome DevTools) \url{https://kritikagarg.github.io/Unnecessary-Archival-Replay-Requests/MREcarousel.html}}
\label{fig:mre_noimg_live}
\end{figure}

Our implementation is similar to the radio commercial example we described in Section \ref{carousel_ex}. We used jQuery \cite{Jquery}, a feature-rich JavaScript library, to create a dynamic carousel. We created a carousel that displays three images every second and generates an HTTP GET request for each image. We hosted this carousel demo using GitHub Pages \cite{githubpages} and then tested it in the Chrome browser to observe its behavior. We made two variations of this demo. In the first variation, the requested image resources are available, and in the second variation, the requested image resources are not available. When the images are available, the browser caches the images received from the first request and then serves the consecutive image requests from the memory cache (Figure~\ref{fig:mre_working}). When the images are not available, the carousel requests the images continuously from the server. Since the requested images are not available, the browser receives HTTP 404 Not Found responses. We noticed that in this scenario the browser does not cache the HTTP 404 response to the first image request (Figure~\ref{fig:mre_noimg_live}). This means continuous requests are made to the web server for unavailable resources. 

\begin{figure}[ht]
\centering
\includegraphics[width=1\textwidth]{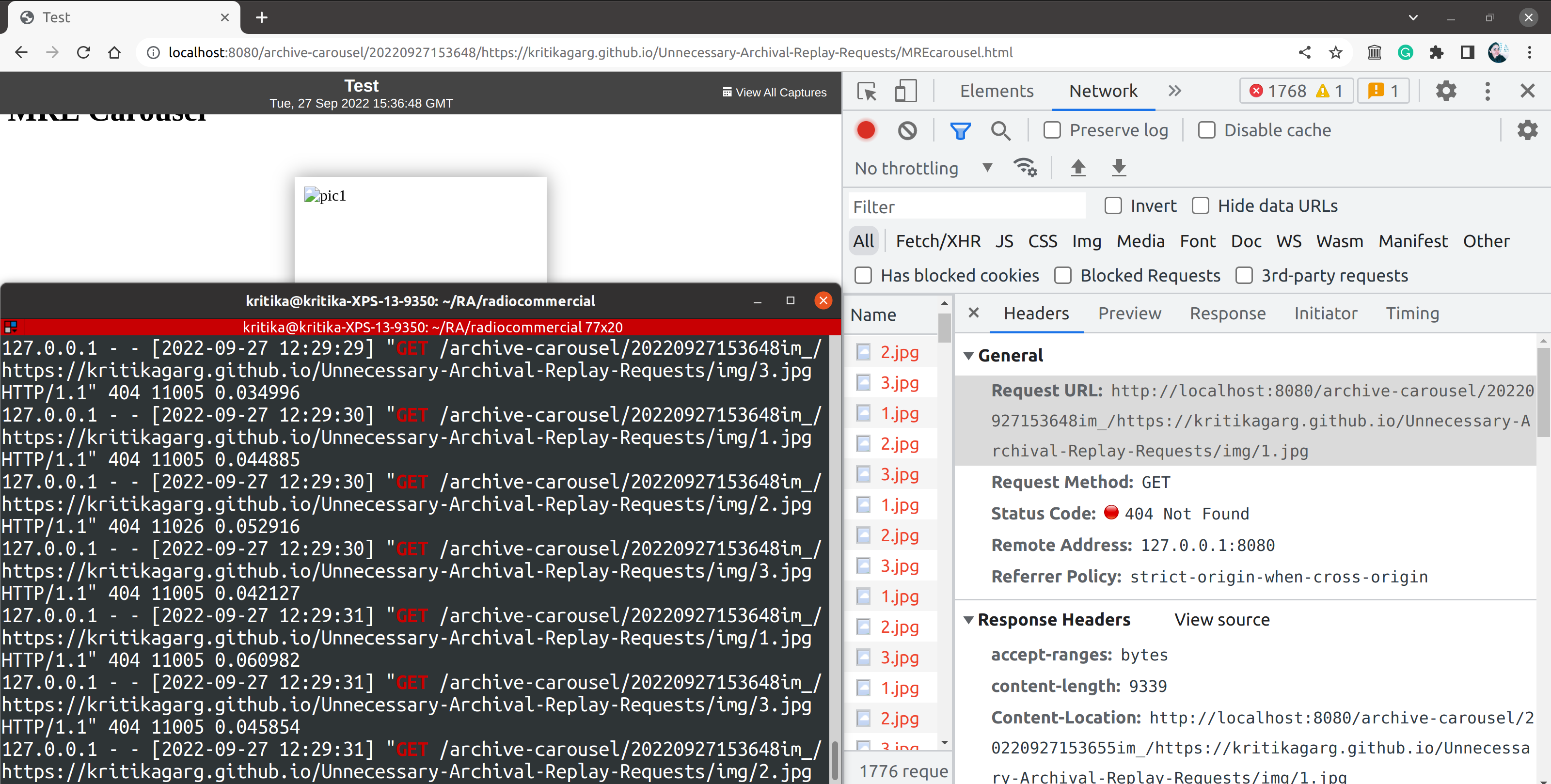}
\caption{Web server logs of pywb (the black terminal screen) showing recurring requests made by the MRE carousel memento}
\label{fig:mre_noimg_pywb}
\end{figure}

We archived this demo page to test its behavior in a web archive environment. We generated a HAR (HTTP Archive) file which tracks all the detailed logging of web browser's HTTP transactions with the demo page. We used the har2warc Python package \cite{har2warc} to convert this HAR file into a WARC (Web Archive) format. We replayed the WARC file locally using pywb, a web archive replay system that allows users to replay archived web content in their browser \cite{pywb}. Figure~\ref{fig:mre_noimg_pywb} shows the archived demo page continuously sending requests to the pywb server for the missing images. The terminal in Figure~\ref{fig:mre_noimg_pywb} shows the web server logs for the requested images.

\section{Approach: Caching HTTP 404 Responses}
\label{solution}
\begin{figure}[ht]
\centering
\includegraphics[width=1\textwidth]{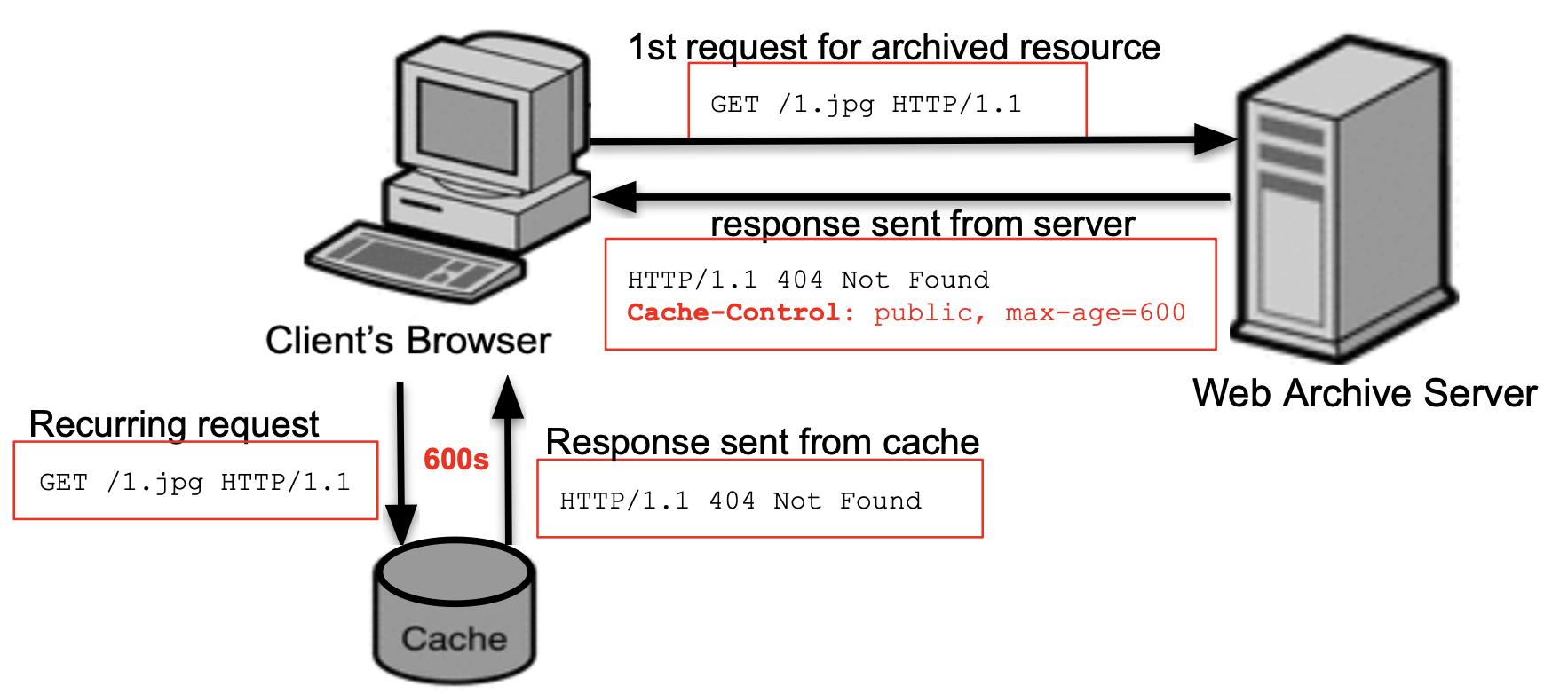}
\caption{A diagram illustrating our proposed solution to avoid recurring requests to archive server by caching HTTP 404 responses}
\label{fig:arch_sol}
\end{figure}

\begin{figure}[htbp]
\centering
\includegraphics[width=1\textwidth]{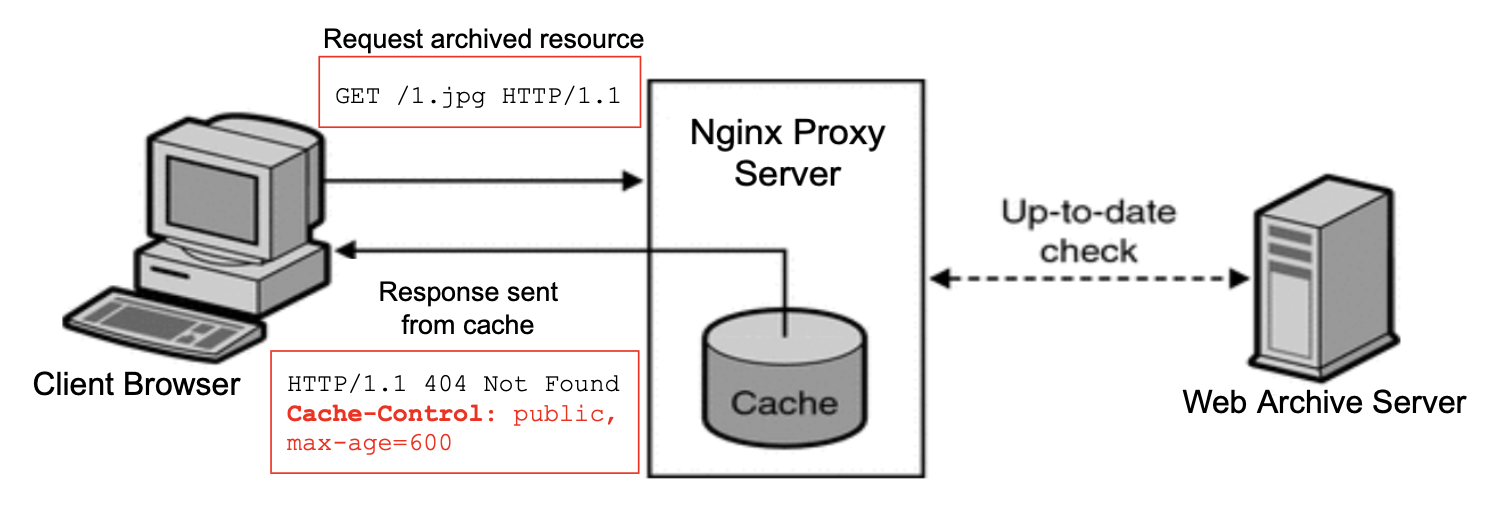}
\caption{Implementing Cache-Control HTTP header for MRE using Nginx as a reverse proxy server to control the network traffic between the browser and the pywb}
\label{fig:arch_sol_implement}
\end{figure}

After studying the behavior of our demo carousel, we understood that we could avoid recurring requests without being obtrusive to the web archives by caching HTTP 404 responses. Figure~\ref{fig:arch_sol} shows our proposed solution where recurring requests are served via memory cache for desired amount of time. We set up Nginx \cite{Nginx} as a reverse proxy server to control the network traffic between a client and the archive server. This allowed us to intercept responses headed for the server and enable Cache-Control for them. Cache-Control is an HTTP header consisting of a set of directives that define when/how a response should be cached and the response’s maximum age before expiring. We configured our Nginx proxy server to add the Cache-Control HTTP header to all outgoing responses (Figure~\ref{fig:arch_sol_implement}), which means responses other than HTTP 404 would also be cached. Web archives generally have a cache mechanism to cache successful responses. However, HTTP 404 error responses are not cached because if the missing resource becomes available later, it could be served immediately to the client. Replaying mementos with missing resources can trigger wasteful network requests to the archive. Our simplified and practical method of caching HTTP 404 responses might cause a brief delay in the time between archiving and serving the missing resource. However, it will help web archives reduce the unnecessary overload on their system caused by these mementos without disrupting the application.

\begin{figure}[ht]
\centering
\includegraphics[width=1\textwidth]{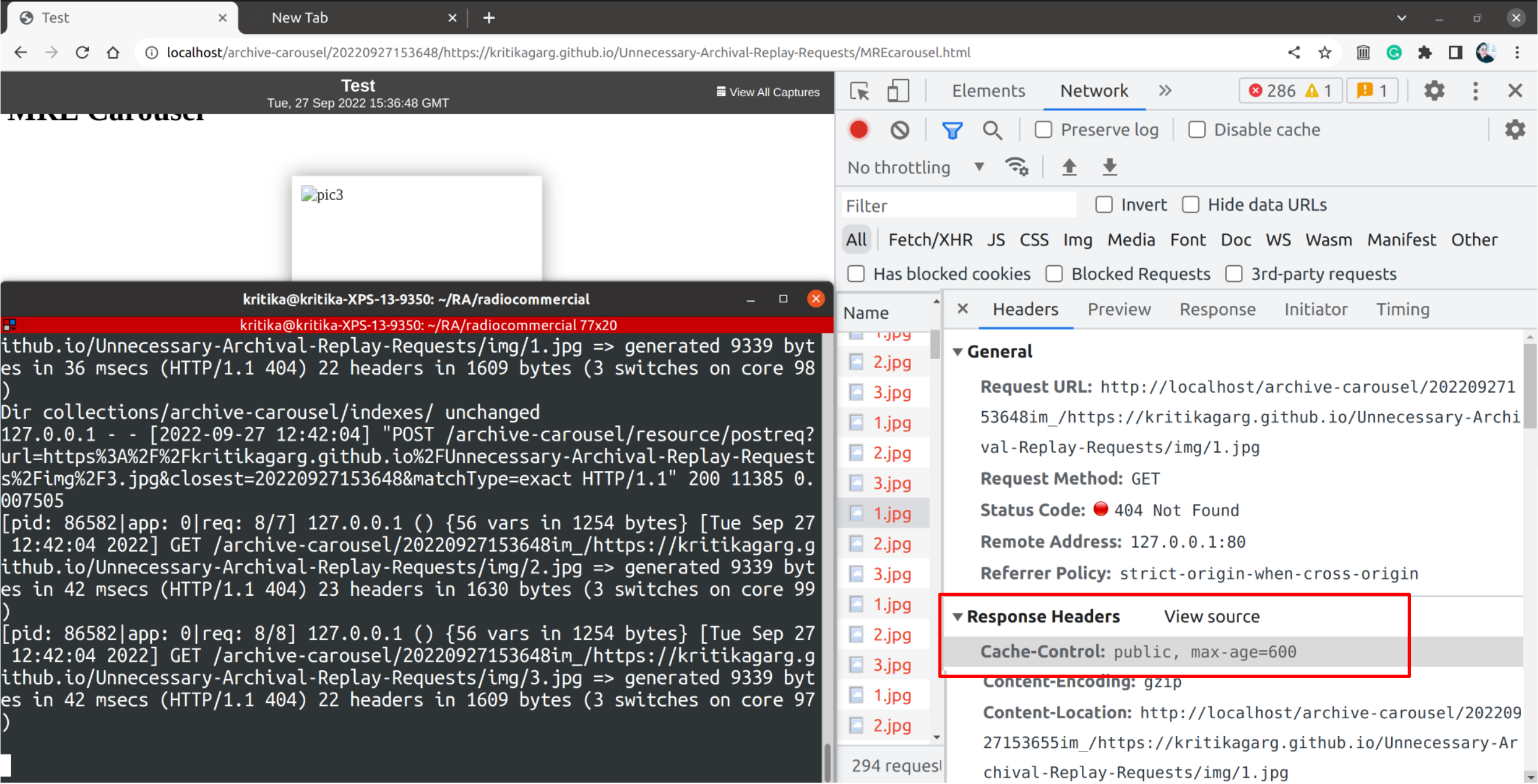}
\caption{Web server logs (the black terminal screen) showing no recurring requests after setting the HTTP Cache-Control response header (request in Chrome DevTools)}
\label{fig:mre_noimg_cache}
\end{figure}

We learned that to implement this solution successfully and set the cache control to all the outgoing requests, we were required to run the pywb with the uWSGI server application  \cite{uswgi}. We used the uwsgi.ini script provided by pywb to launch the server with uwsgi. uWSGI is often used for serving Python-based applications such as pywb in conjunction with the Nginx web server. Nginx offers direct support for the native uwsgi protocol used by the uWSGI application for communication with other servers. We launched the pywb with uWSGI on port 8081 while we ran the Nginx proxy server on port 80 with the configuration shown in Figure~\ref{fig:mre1_sol_config}. 
We replayed the demo carousel on localhost:80 to test its behavior with the Cache-Control HTTP response header in place. In this case, we have set the Cache-Control response header to public. The public response directive indicates that the response can be stored in a shared cache that exists between the origin server and clients. We have also set the max-age directive to indicate the length of time a response is considered fresh, in this case, 600 seconds (10 minutes).\footnote{The max-age value could be adjusted to reflect an archive's accession frequency, the existence of patching / Save Page Now functionality, and other archive-specific preferences.} We observed the effect of the change in the web server logs after setting the Cache-Control response header (Figure~\ref{fig:mre_noimg_cache}). The web server logs did not display any recurring requests because the HTTP 404 responses were cached. Arquivo.pt has implemented our proposed solution. They have added a Cache-Control HTTP response header to cache HTTP 404 responses.

\begin{figure}[ht]
\centering
\begin{lstlisting}[numbers=none, backgroundcolor = \color{white}]
/etc/nginx/sites-available$ cat custom_server.conf 
server {
  listen 80;
  location /static {alias /home/kritika/RA/radiocommercial/static;}
  
  location / {
    uwsgi\_pass localhost:8081; 
    include uwsgi_params;
    uwsgi_param UWSGI_SCHEME $scheme;

    proxy_pass_request_headers   on;
	<@\textcolor{red}{add\_header Cache-Control ``public, max-age=600'' always;}@> 
  }
}
\end{lstlisting}
 \caption{Nginx configuration file to set the Cache-Control HTTP header}
\label{fig:mre1_sol_config}
\end{figure}

\section{Evaluating Rate of Recurring Requests}
\label{rate}
\begin{figure}[!htbp]
    \centering
    \begin{subfigure}[b]{1\textwidth}
        \centering
        \includegraphics[width=1\linewidth]{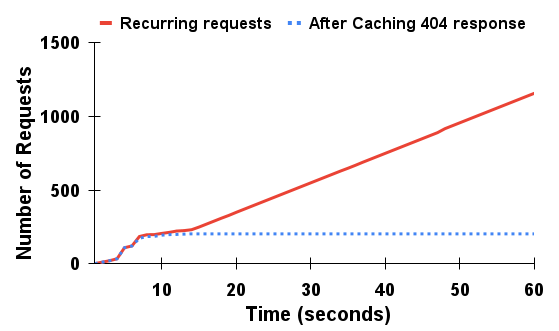}
        \caption{The cumulative number of requests made per second by \texttt{radiocomercial.iol.pt} memento and the anticipated number of requests after introducing the Cache-Control HTTP response header}
        \label{fig:eval_radiopage}
    \end{subfigure}
    
    \begin{subfigure}[b]{1\textwidth}
        \centering
        \includegraphics[width=1\linewidth]{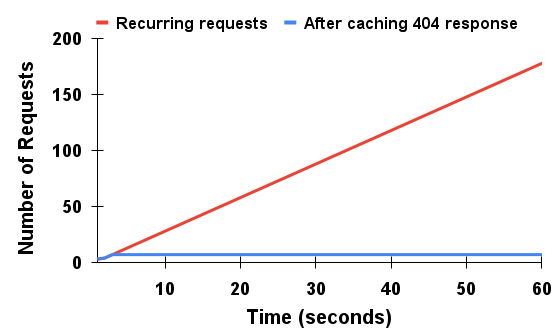}
        \caption{The cumulative number of requests made per second by MRE memento before and after introducing the Cache-Control HTTP response header}
        \label{fig:eval_demopage}
    \end{subfigure}
    \caption{The cumulative number of requests made per second before and after introducing the Cache-Control HTTP response header}
    \label{fig:eval}
\end{figure}

To evaluate the rate of the recurring requests, we first looked at the memento of \texttt{radiocomercial.iol.pt} described in Section \ref{carousel_ex}. We evaluated the number of requests made by the memento before and after introducing the Cache-Control HTTP header. We replayed the memento and recorded the HTTP session using Chrome DevTools. We downloaded this session as a HAR file and analyzed it using haralyzer \cite{haralyzer}. We obtained the number of requests the memento made to the server every second and plotted the cumulative number of requests over time as shown in Figure~\ref{fig:eval_radiopage}. The x-axis represents the time in seconds, and the y-axis represents the cumulative sum of the number of requests. The slope of the line indicates how many new requests are issued every second. The red line demonstrates the wasteful recurring requests. The red line becomes linear after the first 13 seconds. We measured that the memento made 1098.36 requests every minute on average to the web archive. The memento made 203 requests in the first 13 seconds for the essential resources required to replay the memento, and the rest were wasteful requests made for the missing resources. The blue line demonstrates the scenario if the Cache-Control HTTP header were in place. Since we do not control the web archive, we cannot control which headers are returned. After the primary 203 requests for the required resources, there would be no further new requests with the responses being cached.  This anticipated behavior is shown in the figure with the blue line becoming flat.

Figure~\ref{fig:eval_demopage} shows the number of requests made by the memento of MRE described in Section \ref{abstract_model}. In this case, because we control the server, we implemented the HTTP Cache-Control header on responses. We obtained the number of requests our MRE memento made to the server before and after introducing the Cache-Control HTTP header. We found that our example memento’s growth matches the projected behavior in Figure~\ref{fig:eval_radiopage}. We measured that our MRE memento made 174 requests every minute on average to the server. The red line is linear after the first 3 seconds (or first seven requests) due to recurring requests. The blue line shows how this linear growth changed into a  flat straight line after caching, demonstrating that we successfully eliminated the unnecessary recurring requests.

\section{Future Work}
For future work, we propose to look at other examples of frequent requests, such as where the URL changes with an increment variable or a timestamp in the query parameter. For example, appending random query strings to each requested resource would make each request unique. Most web archives do not have a generalizable method to recognize and ignore these query strings, which leads to HTTP 404 responses for the requests for the mementos of these URI-Rs.
One potential solution could be for archival replay systems to have more strategic canonicalization \cite{surt,alam:dissertation,jcdl-2019:alam:mementomap} in place to eliminate such requests. For example, pywb performs fuzzy matching\footnote{\url{https://github.com/webrecorder/pywb/blob/main/pywb/warcserver/index/fuzzymatcher.py}} on the query strings, where it ignores the query parameters and content is loaded from the URL under the Content-Location header. This problem cannot be mitigated by caching because although the server may recognize these multiple URLs to be for the same resource, the client would see them as different URLs. So any caching header applied on one request will not be applied to another from the client's perspective. We propose future work to explore the possibility of adding intelligence in client-side replay libraries like reconstructive \cite{jcdl-2017-alam-reconstructive,reconstructive:gh} and wombat.js \cite{wombat}. These client-side libraries could watch the requests and limit them if they detect any patterns of repetition or similarity in the URLs or the responses they are receiving from the archive. They could serve a prior response of one of the requests to the client using service workers so the request will not go to the web archive server.

\section{Conclusions}
Replaying an archived web page should not cause hundreds or thousands of recurring requests per minute to web archives. 
In this paper, we described various forms in which web archival replay can generate wasteful requests using example web pages archived by Arquivo.pt and the Internet Archive. We provided examples of web pages with banners, carousels, playlists, and web pages that request regular updates (e.g., updates for sports scores, stock prices, and new tweets). We identified that JavaScript triggers recurring HTTP GET requests for the same URL of the missing embedded resources upon replaying the memento. Web archives that try to patch these missing embedded resources from the live web may cause even more unnecessary traffic to the web archive. On a large scale, excessive web traffic could lead to the denial of archival services. We presented that web archives can mitigate unnecessary requests by sending a Cache-Control header on the HTTP 404 responses. We demonstrated this simplified and effective method on a minimal reproducible example memento that initially made 174 requests per minute. After introducing a Cache-Control response header, the memento only made seven requests, eliminating the unnecessary recurring requests.

\section{Acknowledgements}
We are grateful to Daniel Gomes and Fernando Melo of Arquivo.pt for sharing access log data from the Arquivo.pt web archive with us.
\bibliographystyle{splncs04}
\bibliography{main}
\end{document}